\documentclass{article}

\usepackage{graphicx}
\usepackage[left=1.25in,right=1.25in]{geometry}
\usepackage{graphicx}
\usepackage{amsmath}
\usepackage{amssymb}
\usepackage{natbib}
\bibpunct[,]{(}{)}{;}{a}{}{,}  
\usepackage{subfig} 
\usepackage[hyphens]{url}
\usepackage{hyperref}
\hypersetup{colorlinks=true,%
            linkcolor=black,%
            citecolor=black,%
            filecolor=black,%
            urlcolor=black,%
            linktoc=all
}
\usepackage{breakurl}

\newcommand{\bfx}{\mbox{\boldmath{$x$}}}
\newcommand{\bfX}{\mbox{\boldmath{$X$}}}
\newcommand{\bfv}{\mbox{\boldmath{$v$}}}
\newcommand{\bfV}{\mbox{\boldmath{$V$}}}

\newcommand{\bfomega}{\mbox{\boldmath{$\omega$}}}
\newcommand{\bfalpha}{\mbox{\boldmath{$\alpha$}}}
\newcommand{\bfbeta}{\mbox{\boldmath{$\beta$}}}
\newcommand{\bfgamma}{\mbox{\boldmath{$\gamma$}}}
\newcommand{\bfrho}{\mbox{\boldmath{$\rho$}}}
\newcommand{\bftau}{\mbox{\boldmath{$\tau$}}}
\newcommand{\bftheta}{\mbox{\boldmath{$\theta$}}}
\newcommand{\bfvarphi}{\mbox{\boldmath{$\varphi$}}}

\newcommand{\bfxi}{\mbox{\boldmath{$\xi$}}}
\newcommand{\bfzeta}{\mbox{\boldmath{$\zeta$}}}

\newcommand{\tT}{\mbox{\tiny{$T$}}}

\begin{document}

\title{A Flexible Partially Linear Single Index Proportional Hazards Regression Model for Multivariate Survival Data
}

\author{Na Lei  (na@wolters-lei.com)       \and
        Mark A. Wolters (mark@mwolters.com) \and
        Wenqing He (whe@stats.uwo.ca)
}
\date{February 28, 2019}

\maketitle

\begin{abstract}
    We address the problem of survival regression modelling with multivariate responses and nonlinear covariate effects. Our model extends the proportional hazards model by introducing several weakly-parametric elements: the marginal baseline hazard functions are expressed as piecewise constants, association is modelled with copulas, and nonlinear covariate effects are handled by a single-index structure using a spline.  The model permits a full likelihood approach to inference, making it possible to obtain individual-level survival or hazard function estimates.  Performance of the new model is evaluated through simulation studies and application to the Busselton health study data. The results suggest that the proposed method can capture nonlinear covariate effects well, and that there is benefit to modeling the association between the correlated responses.  
\end{abstract}

\section{Introduction}\label{intro}

Multivariate survival data arise frequently in health and medical studies. Examples include epidemiological cohort studies involving relatives, in which the age at disease occurrence is collected from members of the same family; and clinical trials, in which multiple event times are recorded for each individual. In these cases  survival times are correlated, and it is not appropriate to model them as independent events. To extract scientifically useful information from such data, it is appropriate to use multivariate, rather than univariate, survival analysis techniques.   

Two broad categories of multivariate survival data can arise.  If the survival times are observed in some specified order, it is referred to as the longitudinal or sequential setting.  If the observations arise from different diseases in the same individual, or the same disease in related individuals, this is the parallel setting. In the parallel setting, each set of correlated observations is called a \emph{cluster}. 

The present work considers multivariate extensions of the proportional hazards (PH) model \citep{Cox1972}. Though our approach can be applied to any  type of multivariate survival data, we focus on the parallel setting.  

In the univariate case, the PH model can be generalized to have the form
$$\lambda(t|\bfx)=\lambda_0(t)r(h(\bfx)),$$
where $\lambda(t|\bfx)$ is the conditional hazard  function given the covariates, $\lambda_0(t)$ is the baseline hazard, and $r(\cdot)$ is a positive function. Conventionally, $r(\cdot)$ is the exponential function and $h(\bfx)=\bfbeta^{\tT}\bfx$, where $\bfx$ and $\bfbeta$ are vectors of covariates and coefficients, respectively. When $\lambda_0(t)$ is unspecified, it is referred to as the Cox PH model, and inference is done by a partial likelihood approach \citep{Cox1975}. When $\lambda_0(t)$ is assumed to follow a specific distribution, it is called a parametric PH model and the full likelihood can be used for inference.

The conventional Cox PH model assumes the covariates have a linear effect on the log hazards ratio. Many nonparametric approaches have been proposed to relax this assumption \citep[e.g.,][]{Tibshirani1987,Fan1997,Kooperberg1995}. While these approaches improve the flexibility of the model, they become limited by the curse of dimensionality as the number of predictors grows.  Additive models \citep{Hastie1990} and single index models \citep{Hardle1993} can ameliorate this problem.  

The single index approach  generalizes the linear predictor $\bfbeta^{\tT}\bfx$ into $\psi(\bfbeta^{\tT}\bfx)$, where $\psi$ is an unknown smooth function of one variable. In a \emph{partially linear} single index model, the overall covariate effect takes the form $\bfbeta^{\tT}\bfx+\psi(\bfalpha^{\tT}\bfv)$. Some covariates ($\bfx$) enter the model linearly, while others ($\bfv$) enter nonlinearly through the single index. This approach has been used to extend the generalized linear model \citep{Carroll1997}. In the  survival context, \cite{Lu2006} used this approach with kernel methods to estimate $\psi$ and profile quasi-likelihood methods of inference. \cite{Sun2008} also applied a partially linear single index PH model, using a spline to approximate the smooth function.

When turning from univariate to multivariate observations, the problem of modelling association between responses arises. Three common approaches are marginal models \citep{WS88SiM,LSC93JRSSB}, frailty models \citep{Vaupel1979,Clayton1978} and copula models \citep{F79AM,G87B,Joe97}. In survival modelling, a number of authors have considered the marginal model approach \citep{Yu2008,Huster1989,Wei1989,Lee1992,Lin1994}, and the frailty approach is described by \cite{Hougaard2000} and \cite{Lawless2003}.  \cite{He2003} used the copula approach, with piecewise constants or splines used to represent the baseline hazard function in both marginal and conditional PH models. 

The model proposed here combines many of the above ideas.  We develop a multivariate partially linear single index PH model with a copula model for the association, a piecewise constant baseline hazard function, and a spline for the single index function.  The model contains highly shape-flexible elements usually described as nonparametric, yet it can be expressed with a moderate number of parameters and, because it is fully specified, can be estimated using standard maximum likelihood methods. Hence we refer to the model as \emph{weakly parametric}.  

The details of the proposed model are provided in Section \ref{Sec:PHmodel}.  Estimation and inference questions are addressed in Section \ref{Sec:PHparameters}.  The performance of the model is  explored  through simulation in Section \ref{Sec:PHsimulation}, and by application to a real data set in Section \ref{Sec:Application}.  Some conclusions about the proposed model are summarized at the end. 

\section{The Proposed Model}\label{Sec:PHmodel}

In this section we describe the various components of the model in detail, leading up to an explicit statement of the likelihood in Section \ref{ssec:likelihood}.    

\subsection{Notation}\label{ssec:Notation}

Let $T_{ij}$ and $C_{ij}$ be the failure and censoring times of the $j$th observation in the $i$th cluster ($j=1,\ldots,m_i$, $i=1,\ldots,n$). The observed data $(y_{ij},\bfx_{ij},\bfv_{ij},\delta_{ij})$ are realizations of the variables $(Y_{ij},\bfX_{ij},\bfV_{ij},\Delta_{ij})$, where $Y_{ij}=\text{min}(T_{ij},C_{ij})$ is the observed event time, and $\Delta_{ij}=I(T_{ij}\leq C_{ij})$ is the censoring indicator. The covariates $\bfX_{ij}$ and $\bfV_{ij}$ are assumed to have linear and nonlinear effects, respectively. The survival times within each cluster are assumed correlated, while the observations from different clusters are assumed independent. For ease of exposition and without loss of generality, we restrict our attention to the case of bivariate responses (clusters of size two).

For an observation in the $i$th cluster, the proposed marginal hazard function is
$$\lambda_{ij}(t|\bfx_{ij},\bfv_{ij})=\lambda_{0j}(t)\text{exp}\{\bfbeta^{\tT}\bfx_{ij}+\psi(\bfalpha^{\tT}\bfv_{ij})\},~~~j=1,~2,$$
where $\bfbeta$ is a $p$-vector of coefficients, $\bfalpha$ is a $q$-vector of coefficients, $\psi(\cdot)$ is a smooth function, and $\psi(\bfalpha^{\tT}\bfv_{ij})$ is the single index structure.  Note that the baseline hazard function $\lambda_0$ is indexed by $j$, while $\bfbeta$ and $\bfalpha$ are not.  This means that the baseline hazard is allowed to differ among members of the same cluster, while the covariate effects are not.  This is a modelling choice made to limit the number of parameters in our model.  Extending the model to allow different coefficients for the two variates poses no difficulties, as estimation and inference procedures are unaffected. 

Let $f(t_1,t_2)$ denote the joint density function of the failure times. The likelihood contributed from the $i$th cluster is
\begin{eqnarray}
\label{eq:Li}
L_i&=&f(t_{i1},t_{i2})^{\delta_{i1}\delta_{i2}}\Bigg[\frac{-\partial S(t_{i1},t_{i2})}{\partial t_{i1}}\Bigg]^{\delta_{i1}(1-\delta_{i2})}
\nonumber \\
&&\times\Bigg[\frac{-\partial S(t_{i1},t_{i2})}{\partial t_{i2}}\Bigg]^{(1-\delta_{i1})\delta_{i2}}S(t_{i1},t_{i2})^{(1-\delta_{i1})(1-\delta_{i2})},
\end{eqnarray}
as shown in \citet{Lawless2003}.  This likelihood is for clustered data.  The sequential-data likelihood is the same, except the third factor on the right hand side becomes unity, because $(\delta_{i1},\delta_{i2}) = (0,1)$ can never be observed.

To draw inferences from the model, this likelihood must be expressed as a function  of its parameters. To do this, we must specify the mathematical form of the baseline hazard functions $\lambda_{0j}(t_j)$ and the smooth function $\psi(\cdot)$, as well as the manner in which the joint survival function $S(t_{i1},t_{i2})$ is obtained from the marginal survival functions. We propose to use a piecewise constant to model the baseline hazard functions \citep{He2003}, a spline for the smooth function $\psi(\cdot)$, and the Clayton copula model \citep{Clayton1978} for the joint survival function.

\subsection{Baseline hazard function}\label{Sec:piecewise constant}

A piecewise constant approximation to the baseline hazard relaxes the assumption in the parametric PH model to allow greater shape flexibility.  Assume that the marginal baseline hazard functions $\lambda_{0j}(t_{j})$, $j=1,2,$ have piecewise constant forms as follows:
$$\lambda_{01}({t})=\rho_k, ~~\text{where}~{t}\in A_k=(a_{k-1}, a_k], ~k=1,\ldots,r,$$
and
$$\lambda_{02}({t})=\tau_l, ~~\text{where}~{t}\in B_l=(b_{l-1}, b_l], ~l=1,\ldots,s,$$
where $0=a_0<a_1<\ldots<a_r=\infty$ and $0=b_0<b_1<\ldots<b_s=\infty$ are pre-chosen sequences of constants, also called cut points, and $\bfrho=(\rho_1,\ldots,\rho_r)^{\tT}$ and $\bftau=(\tau_1,\ldots,\tau_s)^{\tT}$ are unknown positive constants to be estimated. $A_k$ and $B_l$ are the intervals defined by the sequence of cut points. 

The corresponding marginal cumulative hazard functions, obtained by integration over the piecewise constants, are
\begin{eqnarray*}
\Lambda_{01}({t})&=&\sum_{k=1}^r\rho_ku_k({t}) \nonumber \\
\Lambda_{02}({t})&=&\sum_{l=1}^s\tau_lw_l({t}),
\end{eqnarray*}
where $u_k({t})=\max(0,\allowbreak\min(a_k,{t})-a_{k-1})$ and $w_l({t})={\max}(0,{\min}(b_l,{t})-b_{l-1})$  are the lengths of the intersection of the interval $(0,{t})$ with $A_k$ and $B_l$, respectively.

An important question is how to choose the cut points. The choice can be made based on prior assumptions about the marginal distributions of the survival time, or one can start from a small number of cut points, say 2 or 3, and increase the number to observe their effect. It is suggested four or five interior cut points is usually good enough \citep{Lawless1998}. It is desirable to keep roughly the same number of failures within each interval. A practical approach, particularly when the censoring rate is high, is to construct the Kaplan-Meier survival estimate, ignoring covariate effects, and then chose cut points that give roughly the same survival probability in each interval. This is the approach we use in our simulation studies and real data analysis.

\subsection{Single index function}\label{Sec:PH spline functions}

In the proposed model we express the smooth function $\psi(\cdot)$ as a spline.  Our spline is subject to the constraint that $\psi(0)=0$, to ensure that the hazard equals the baseline hazard when both $\bfx$ and $\bfv$ are $\mathbf{0}$.  This constraint can be elegantly handled using the M-spline and I-spline bases (see \cite{Ramsay1988} for definitions and computational details of using these bases).  

A spline of order $m$, with $d$ basis functions, is defined on an interval $[L,U]$, with $m$ knots at each of the boundary points, and $d-m$ interior knots. Denote the $j$th M-spline and I-spline basis functions by $M_j(\cdot)$ and $I_j(\cdot)$, respectively. 

We represent the derivative of the single index function as an M-spline,
\begin{equation*}
\psi'(\bfalpha^{\tT}\bfv)=\sum_{j=1}^d\gamma_jM_j(\bfalpha^{\tT}\bfv)=\bfgamma^{\tT} M(\bfalpha^{\tT}\bfv),
\end{equation*}
where $\gamma_j$ is the coefficient for the $j$th basis function, and $M(u)=(M_1(u),\ldots, M_d(u))^{\tT}$. 

By definition, the I-spline basis functions are the integration of the  M-spline bases. Using this fact, and the requirement that $\psi(0)=0$, the single index function can be expressed as:
\begin{eqnarray*}\label{eq:I-spline}
\psi(\bfalpha^{\tT}\bfv)
&=&\int_0^{\bfalpha^{\tT}\bfv}\psi'(x)dx\\
&=&\sum_{j=1}^d\gamma_j\int_0^{\bfalpha^{\tT}\bfv}M_j(x)dx\\
&=&\sum_{j=1}^d\gamma_j\int_L^{\bfalpha^{\tT}\bfv}M_j(x)dx-\sum_{j=1}^d\gamma_j\int_L^{0}M_j(x)dx\\
&=&\sum_{j=1}^d\gamma_j\left[I_j(\bfalpha^{\tT}\bfv)-I_j(0)\right]\\
&=&\bfgamma^{\tT}\left[I(\bfalpha^{\tT}\bfv)-I(0)\right],
\end{eqnarray*}
where $I(u)=(I_1(u),\ldots, I_d(u))^{\tT}$.

Two additional constraints are required to ensure identifiability of both $\bfgamma$ and $\bfalpha$. The first is that $\parallel\bfalpha\parallel=1$, to prevent solutions of the form $(\bfalpha,\bfgamma)$ and $(c\bfalpha, \bfgamma/c)$ from being equivalent. The second is that one specific element of $\bfalpha$ (for example, its last element) must be positive. This prevents the two solutions $(\bfalpha, \bfgamma)$ and $(-\bfalpha, -\bfgamma)$ from being equivalent. With these two constraints, it is possible to find unique estimates for $\bfgamma$ and $\bfalpha$. 

Regarding the strategy of choosing the number of internal knots, we follow \citet{Lawless1998}, who suggest dividing the range of the function into 4-10 intervals. In our work, we use  4 intervals (3 internal knots) in our spline functions.  Knots are chosen such that each interval
contains roughly the same number of data points.

\subsection{Copula}\label{ssec:Copula}

Development of the joint survival function can be done through the specification of a copula function and the marginal distributions. For clusters of size two, the copula function is a bivariate function $C(u,v;\phi)$, defined on the unit square, where $\phi$ is a parameter that controls the association structure.   The characteristics of a valid copula function \citep[see, e.g.,][]{Shemyakin2006} ensure that when its two arguments are univariate survival functions $S_1$ and $S_2$, the function $C(S_1(t_1),S_2(t_2);\phi)$ is a bivariate survival function having $S_1$ and $S_2$ as its marginal survival functions. The marginal survival functions can have parametric or semiparametric forms.

In our model we use the family of copula models introduced by \citet{Clayton1978}. It gives rise to joint survival functions of the form 
\begin{equation}
\label{eq:Clayton}
S(t_{1},t_{2})=\Big[S_{1}(t_{1})^{-\phi^{-1}}+S_{2}(t_{2})^{-\phi^{-1}}-1\Big]^{-\phi},
\end{equation}
where $\phi\in(0,\infty)$. Larger values of $\phi$ represent weaker association between the two survival times, with $\phi=\infty$ corresponding to independence. The range of $\phi$ can be extended to -1 to accommodate some negative association \citep{Lawless2003}. The Clayton copula can also be formulated for clusters larger than two, still with only a single association parameter $\phi$.

\subsection{The Likelihood}\label{ssec:likelihood}

Once the structure of the baseline hazards, the joint survival function, and the single index term are specified, it is possible to express the likelihood explicitly as a function of the model parameters. 

Using the relationship between the survival function and the hazard function, the marginal survival functions can be written as 
\begin{eqnarray*}
&&S_{i1}({t})=\text{exp}\Big[-\Lambda_{01}({t})\text{exp}\{\bfbeta^{\tT}\bfx_{i1}+\psi(\bfalpha^{\tT}\bfv_{i1})\}\Big],\\
&&S_{i2}({t})=\text{exp}\Big[-\Lambda_{02}({t})\text{exp}\{\bfbeta^{\tT}\bfx_{i2}+\psi(\bfalpha^{\tT}\bfv_{i2})\}\Big],
\end{eqnarray*}
where $\Lambda_{01}({t})=\sum_{k=1}^r\rho_ku_k({t})$, $\Lambda_{02}({t})=\sum_{l=s}^r\tau_lw_l({t})$, and $\psi(\bfalpha^{\tT}\bfv_{ij})=\bfgamma^{\tT}\left[I(\bfalpha^{\tT}\bfv_{ij})-I(0)\right]$. Note that while $S_{ij}(t)$ would more properly be written $S_{ij}(t|\bfx_{ij},\bfv_{ij})$, we have suppressed the dependence on the covariates to ease the notation.

The log likelihood is $l = \sum_{i=1}^n\text{log}L_i$, where $L_i$ is defined in Equation (\ref{eq:Li}). If the Clayton bivariate model is used, the joint survival function has the form given in Equation (\ref{eq:Clayton}), and the terms $\frac{-\partial S(t_{i1},t_{i2})}{\partial t_{i1}}$, $\frac{-\partial S(t_{i1},t_{i2})}{\partial t_{i2}}$, and $f(t_{i1},t_{i2})$ in the log likelihood are:
\begin{eqnarray*}
\frac{-\partial S(t_{i1},t_{i2})}{\partial t_{ij}}&=&
\Big[S_{i1}(t_{i1})^{-\phi^{-1}}+S_{i2}(t_{i2})^{-\phi^{-1}}-1\Big]^{-\phi-1}\Big[S_{ij}(t_{ij})^{-\phi^{-1}}\Big]
\nonumber \\
&&\times \text{exp}\{\bfbeta^{\tT}\bfx_{ij}+\psi(\bfalpha^{\tT}\bfv_{ij})\}\frac{\partial [\Lambda_{0j}(t_{ij})]}{\partial t_{ij}}, ~~~~~j=1,2
\end{eqnarray*}
and
\begin{eqnarray*}
f(t_{i1},t_{i2})&=&\frac{\partial^2 S(t_{i1},t_{i2})}{\partial t_{i1}\partial t_{i2}}
\nonumber \\
&=&\Big[S_{i1}(t_{i1})^{-\phi^{-1}}\Big]\Big[S_{i2}(t_{i2})^{-\phi^{-1}}\Big]\text{exp}\{\bfbeta^{\tT}\bfx_{i1}+\psi(\bfalpha^{\tT}\bfv_{i1})\}
\nonumber \\
&&\times \text{exp}\{\bfbeta^{\tT}\bfx_{i2}+\psi(\bfalpha^{\tT}\bfv_{i2})\}
\frac{\partial [\Lambda_{01}(t_{i1})]}{\partial t_{i1}}
\nonumber \\
&&\times \frac{\partial [\Lambda_{02}(t_{i2})]}{\partial t_{i2}} \Big[S_{i1}(t_{i1})^{-\phi^{-1}}+S_{i2}(t_{i2})^{-\phi^{-1}}-1\Big]^{-\phi-2}(1+\phi^{-1}).
\end{eqnarray*}
Each of these expressions contains partial derivatives of the marginal cumulative hazard functions $\Lambda_{01}$ and $\Lambda_{02}$.  These derivatives are simply the appropriate elements of $\bfrho$ and $\bftau$, the parameters of the piecewise-constant marginal baseline hazards. 

\section{Estimation and Inference}\label{Sec:PHparameters}

The likelihood just described involves parameters
$$\bftheta=(\phi,\bfrho^{\tT},\bftau^{\tT},\bfalpha^{\tT},\bfbeta^{\tT},\bfgamma^{\tT})^{\tT},$$
where $\phi$ is the association parameter, and the other subvectors of parameters are $\bfrho=(\rho_1,\ldots,\rho_r)^{\tT}$ and $\bftau=(\tau_1,\ldots,\tau_s)^{\tT}$ for the piecewise constants, $\bfalpha=(\alpha_1,\ldots,\alpha_q)^{\tT}$ for the nonlinear covariates, $\bfbeta=(\beta_1,\ldots,\beta_p)^{\tT}$ for the linear covariates, and $\bfgamma=(\gamma_1,\ldots,\gamma_d)^{\tT}$ for the single index.

The model imposes constraints on some of the parameters.  They are:
$$\rho_1,\ldots,\rho_r>0,$$
$$\tau_1,\ldots,\tau_s>0,$$
$$\phi>0,$$
$$\parallel\bfalpha\parallel=1,~{\rm and}~\alpha_q>0.$$
The constraint that $\psi(0)=0$ is not included in the above list, because it is satisfied by construction using the I-spline, as described in Section \ref{Sec:PH spline functions}.

We now describe how estimation and inference may be carried out, while respecting the constraints on the parameters. 

\subsection{Optimizing the Likelihood}\label{ssec:optimizing}

Several parameter transformations can eliminate the constraints, allowing us to avoid a constrained optimization problem. For the positive parameters $\rho_k$, $\tau_l$, and $\phi$, a log transformation is applied. The transformed parameters are denoted by $\xi_k$, $\zeta_l$ and $\varrho$:
$$\xi_k=\log\rho_k, ~~~k=1,\ldots,r,$$
$$\zeta_l=\log\tau_l, ~~~l=1,\ldots,s,$$
and
$$\varrho=\log\phi.$$

For $\bfalpha$, the transformation takes two steps. First we use a trigonometric transformation on $\bfalpha$ to enforce the norm constraint and the positivity of the $q$th element. Parameter  $\bfomega=(\omega_1,\ldots,\omega_{q-1})^{\tT}$ is introduced, as follows:
\begin{eqnarray}
\label{eq:polar}
\alpha_1&=&\sin\omega_1\sin\omega_2\cdots\sin\omega_{q-2}\sin\omega_{q-1} \nonumber\\
\alpha_2&=&\sin\omega_1\sin\omega_2\cdots\sin\omega_{q-2}\cos\omega_{q-1} \nonumber\\
&\ldots\\
\alpha_{q-1}&=&\sin\omega_1\cos\omega_2 \nonumber \\
\alpha_q&=&\cos\omega_1 \nonumber
\end{eqnarray}
where $\omega_1,\ldots,\omega_{q-1}\in[-\frac{\pi}{2},\frac{\pi}{2}]$.  To eliminate the bound constraint on each $\omega$, define a new parameter $\bfvarphi=(\varphi_1,\ldots, \varphi_{q-1})^{\tT}$, where $\varphi_i = \log\left[(\frac{\pi}{2}+\omega_i)/(\frac{\pi}{2}-\omega_i)\right].$ After these two steps, the relationship between $\bfalpha$ and $\bfvarphi$ is
\begin{eqnarray*}
\alpha_1&=&\sin\Big[\frac{e^{\varphi_1}-1}{e^{\varphi_1}+1}\frac{\pi}{2}\Big]\sin\Big[\frac{e^{\varphi_2}-1}{e^{\varphi_2}+1}\frac{\pi}{2}\Big]\cdots\sin\Big[\frac{e^{\varphi_{q-1}}-1}{e^{\varphi_{q-1}}+1}\frac{\pi}{2}\Big] \nonumber \\
\alpha_2&=&\sin\Big[\frac{e^{\varphi_1}-1}{e^{\varphi_1}+1}\frac{\pi}{2}\Big]\sin\Big[\frac{e^{\varphi_2}-1}{e^{\varphi_2}+1}\frac{\pi}{2}\Big]\cdots\cos\Big[\frac{e^{\varphi_{q-1}}-1}{e^{\varphi_{q-1}}+1}\frac{\pi}{2}\Big] \nonumber \\
&\ldots \nonumber \\
\alpha_q&=&\cos\Big[\frac{e^{\varphi_1}-1}{e^{\varphi_1}+1}\frac{\pi}{2}\Big].
\end{eqnarray*}

With these transformations, the set of unconstrained parameters is$$\bftheta^\ast=(\varrho,\bfxi^{\tT},\bfzeta^{\tT},\bfvarphi^{\tT},\bfbeta^{\tT},\bfgamma^{\tT})^{\tT},$$
where $\bfxi=(\xi_1,\ldots,\xi_r)^{\tT}$, $\bfzeta=(\zeta_1,\ldots,\zeta_s)^{\tT}$, $\bfvarphi=(\varphi_1,\ldots,\varphi_{q-1})^{\tT}$, $\bfbeta=(\beta_1,\ldots,\beta_p)^{\tT}$, and $\bfgamma=(\gamma_1,\ldots,\gamma_d)^{\tT}$. Under this parametrization, the maximum likelihood estimates can be found using a standard optimization algorithm such as the Newton-Raphson method. 

\subsection{Inference}\label{ssec:Inference}

Since we have the full likelihood, standard large-sample maximum likelihood results can be used for statistical inference. The variance estimate of the maximum likelihood estimator $\widehat{\bftheta^\ast}$ can be obtained from the inverse of the observed information matrix. The second derivatives are inconvenient to get, so one can use \citep{He2003}
\begin{equation}
\label{eq:var.theta*}
\widehat{\text{var}}\left(\widehat{\bftheta^\ast}\right)=\Bigg[\sum_{i=1}^n\Big(\frac{\partial {\log}L_i}{\partial \bftheta^\ast}\Big)\cdot\Big(\frac{\partial {\log}L_i}{\partial \bftheta^\ast}\Big)^{\tT}\Bigg]^{-1}_{\bftheta^\ast=\widehat{\bftheta^\ast}}.
\end{equation}
In the numerical demonstrations of the next two sections, we use this formula to obtain the variance estimate of $\widehat{\bftheta^\ast}$.  

Once the variance estimate is found, the delta method can be applied to calculate the variance estimate for $\hat{\bftheta}$, the MLE under the original parametrization. Denote by $G$ the map from $\bftheta^\ast$ to $\bftheta$ (that is, $\bftheta = G(\bftheta^\ast)$).  Then $\hat{\bftheta}$ is asymptotically normal with an estimated asymptotic variance-covariance matrix:
\begin{eqnarray}
\label{eq:var.theta}
\text{var}(\hat{\bftheta})&=&\text{var}\left(G\left(\widehat{\bftheta^\ast}\right)\right) \nonumber \\
&=&G'\left(\widehat{\bftheta^\ast}\right)\text{var}\left(\widehat{\bftheta^\ast}\right)G'\left(\widehat{\bftheta^\ast}\right)^{\tT},
\end{eqnarray}
where $G'$ denotes the gradient of $G$ with respect to $\bftheta^\ast$.
Based on the relationships between $\bftheta$ and $\bftheta^\ast$ given in the previous section, $G'$ can be worked out analytically. To obtain our variance estimator $\widehat{\text{var}}(\hat{\bftheta})$, we substitute $\widehat{\text{var}}(\widehat{\bftheta^\ast})$ for $\text{var}(\widehat{\bftheta^\ast})$ in the above formula.

\section{Simulation Studies}\label{Sec:PHsimulation}

In this section, the performance of the proposed model is assessed through simulation. The data-generating process is described first.  Simulated data was generated with four conditions varied in a full factorial design.  Estimation results from replicates at each design point are used to draw conclusions about the proposed model's performance.  One of the design points was taken as a baseline case, examined in more detail.  At the baseline conditions, a variant of the proposed model with no single index structure was also fit to the data, to assess the benefit of including nonlinear covariate effects. 

\subsection{Experimental conditions}\label{ssec:ExperimentalConditions}

Data were generated as follows. The baseline hazard function was set to be a Weibull hazard with scale parameter 1 and shape parameter $p$. The true marginal hazard functions were set to 
$$\lambda_{ij}(t|x_{ij},\bfv_{ij})=\lambda_{0j}(t)\text{exp}\left\{\beta x_{ij}+3\text{sin}(2\bfalpha^{\tT}\bfv_{ij})\right\},~~j=1,2.$$
The expression $3\text{sin}(2\bfalpha^{\tT}\bfv_{ij})$ is used as the smooth function $\psi(\cdot)$, following \cite{Sun2008}.  Covariates $(x_{ij},\bfv_{ij})$ were generated independently for each member of each cluster, with $x_{ij}$ consisting of a single Bernoulli(0.5) variate, and $\bfv_{ij}$ consisting of three independent $U(-1,1)$ variates.  The nonlinear regression coefficients are $\bfalpha=(1, 1, 1)^{\tT}/\sqrt{3}$ and the linear regression coefficient is $\beta=1$.

Within-cluster association is handled using the Clayton model, with parameter $\phi$.  Bivariate survival times $(t_{i1},t_{i2}),~i=1,\ldots,n$ were generated using inverse transform sampling. Censoring was performed by choosing a fixed censoring time $C$ for all individuals, and adjusting $C$ to obtain the desired censoring rate.

For estimating the proposed model, four pieces were used in both baseline hazard function estimates, and three interior knots were used in the spline estimate of $\psi(\cdot)$.

Simulated data was generated at various combinations of four factors: sample size ($n$), association parameter ($\phi$), Weibull shape parameter ($p$), and censoring rate. We refer to the settings $n=200$, $\phi=0.5$, and $p=1.5$, with 50\% censoring, as the default or baseline experimental scenario.  Two hundred replicate data sets were first generated from this scenario.  The data were fit using the proposed model as well as the \emph{PH linear model}, which has the same structure as the proposed model, but without the single index function.  Its marginal hazard functions have the form 
$$\lambda_{ij}(t|x_{ij},\bfv_{ij})=\lambda_{0j}(t)\text{exp}\{\beta x_{ij}+\tilde{\bfalpha}^{\tT}\bfv_{ij}\},~j=1,2.$$ In the PH linear model, the single index structure is replaced by linear predictor $\tilde{\bfalpha}^{\tT}\bfv_j$. Note that parameter vector $\tilde{\bfalpha}$ is not subject to the same unit norm constraint enforced when the single index is used.  To make the estimates of $\tilde{\bfalpha}$ comparable with the estimates from the proposed model, we factor each estimate as $\hat{\tilde{\bfalpha}}=\hat{b}\hat{\bfalpha}$, where $\parallel\hat{\bfalpha}\parallel=1$ and $\hat{b}$ is interpreted as the slope for a linear regression on $\hat{\bfalpha}^{\tT}\bfv$.

After this, a full factorial experiment was conducted, with the experimental factors taking the following levels: $n \in \{80,200\}$, $\phi \in \{0.5, 1, 4\}$, $p \in \{0.5, 1.5\}$, and censoring rate 20\% or 50\%. The levels chosen for the association parameter cover a range from fairly strong dependence between $T_1$ and $T_2$ (at $\phi=0.5$) to moderate dependence (at $\phi=4$).

\subsection{Performance in the default scenario}\label{ssec:DefaultPerf}

Table \ref{tab:PHpara} gives  estimation results for the regression parameters $\bfalpha$ and $\beta$, and the association parameter $\phi$, for the default set of experimental conditions. Also shown are the transformed single-index coefficients  $\bfvarphi$ and transformed association parameter $\varrho$. In the table, SD($\hat{\cdot}$) is the sample standard deviation of the estimates across replicates, A\{SE($\hat{\cdot}$)\} is the average of the model based standard errors (using Equation \ref{eq:var.theta*}), and Cov.prob.\text{ }is the estimated coverage probability of the $95\%$ confidence intervals. The table shows that the estimates are close to the true values, and the coverage probabilities have values close to the nominal rate.  Only the association parameter estimates have appreciable bias, and they are also more variabe than the other estimates.  Note that A\{SE($\hat{\cdot}$)\} is close to SD($\hat{\cdot}$), suggesting the asymptotic variance estimates are reasonable. This was true for almost all experimental conditions (with  exceptions noted in Section \ref{ssec:OtherCond}), so we only report SD($\hat{\cdot}$) henceforth.  

\begin{table}
\caption{Estimates for parameters $\bfalpha$, $\beta$,  $\phi$, $\bfvarphi$, and $\varrho$ in the default scenario.}
\label{tab:PHpara}
\begin{center}
\begin{tabular}{lccccccccc}
\hline\noalign{\smallskip}
& \multicolumn{5}{c}{Original parametrization} && \multicolumn{3}{c}{Transformed}\\
                           & $\alpha_1$ & $\alpha_2$ & $\alpha_3$ & $\beta$ & $\phi$ && $\varphi_1$ & $\varphi_2$ & $\varrho$\\
\noalign{\smallskip}\hline\noalign{\smallskip}                    
True                       & 0.577      & 0.577      & 0.577      & 1.000   & 0.500  && 1.412       & 1.099       & -0.693    \\
Bias                       & -0.002     & 0.002      & -0.001     & -0.008  & -0.045 && 0.003       & -0.005      & -0.118    \\
SD($\hat{\cdot}$)          & 0.016      & 0.017      & 0.018      & 0.118   & 0.100  && 0.044       & 0.040       & 0.212     \\
A\{SE($\hat{\cdot}$)\}     & 0.017      & 0.017      & 0.017      & 0.128   & 0.109  && 0.042       & 0.043       & 0.235     \\
Cov.prob.                  & 0.950      & 0.950      & 0.930      & 0.965   & 0.850  && 0.930       & 0.950       & 0.910     \\
\noalign{\smallskip}\hline
\end{tabular}
\end{center}
\end{table}

Table \ref{tab:compare} compares estimation results for the PH linear model to the proposed model.  Point estimates of $\bfalpha$ are nearly equal for both models, but the PH linear model has higher variances.  For $\beta$, the PH\ linear model has considerable bias and much larger variance; and for $\phi$, both bias and variance are extreme. 

\begin{table}
\caption{Regression parameter estimates for the PH linear model and the proposed model in the default scenario.}
\label{tab:compare}
\begin{center}
\begin{tabular}{lccccc}
\hline\noalign{\smallskip}
& $\alpha_1$ & $\alpha_2$ & $\alpha_3$ & $\beta$ & $\phi$ \\
\noalign{\smallskip}\hline\noalign{\smallskip}
True & 0.577 & 0.577 & 0.577 & 1.000 & 0.500 \\[1.5mm]
Estimate (proposed) & 0.575 & 0.577 & 0.579 & 0.971 & 0.455 \\
SD($\hat{\cdot}$)(proposed) & 0.017 & 0.018 & 0.017 & 0.100 & 0.128 \\[1.5mm]
Estimate (linear) & 0.571 & 0.576 & 0.581 & 0.709 & 3.060 \\
SD($\hat{\cdot}$) (linear) & 0.041 & 0.042 & 0.041 & 0.178 & 2.226\\
\noalign{\smallskip}\hline
\end{tabular}
\end{center}
\end{table}

\begin{figure*}
\centering
\subfloat[$\Lambda_{01}$]{\label{fig:Lambda0Li}\includegraphics[width=0.475\textwidth]{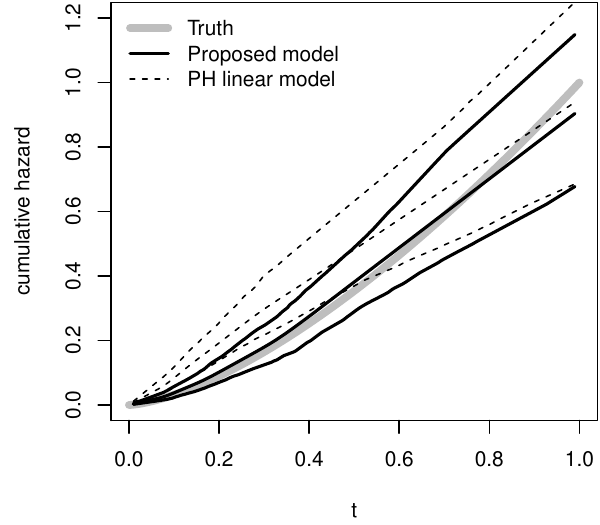}}
~~~\subfloat[$\psi(\cdot)$]{\label{fig:psiLi}\includegraphics[width=0.475\textwidth]{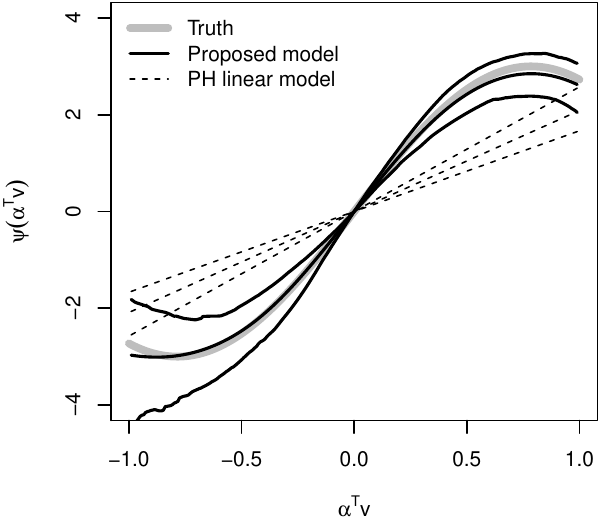}}
\caption{Results for estimation of the cumulative baseline hazard function ($\Lambda_{01}$) and single index function ($\psi$), for 200 replicates at the baseline experimental conditions.  For both the proposed model and the PH linear model, three lines are shown: the middle line is the average estimate, and the bounding lines are the pointwise 5th and  95th percentiles. In panel (b), recall that $\psi(0) = 0$ by constraint.}
\label{fig:Lambda0+psi+Linear}
\end{figure*}

The poor suitability of the PH linear model for the simulation data is more clearly seen in Figure \ref{fig:Lambda0+psi+Linear}, which shows estimation performance for the cumulative baseline hazard function $\Lambda_{01}$ and the nonlinear smooth function $\psi(\cdot)$. The linear model fails to recognize the curvature of either function, yielding very poor estimates.  As can occur with a misspecified model, not only is there large bias, but the variance is too small to indicate a serious problem.  The proposed model, by contrast, provides a nearly unbiased estimate with a similar level of variability.

\subsection{Performance at other conditions}\label{ssec:OtherCond}

Table \ref{tab:othertable} summarizes the estimation quality of the proposed model at the full set of  conditions in the factorial experiment. For brevity, and because they are the parameters of most importance, the table only includes results for regression parameters $\bfalpha$ and $\beta$, along with association parameter $\phi$. The table includes sub-tables for the two values of $p$. 

The table shows that increasing the sample size from 80 to 200 has the expected effect: bias decreases and estimates become more precise. Similarly, increasing the censoring rate from 20\% to 50\% has the expected effect of increasing the standard errors. 

Estimation of the nonlinear regression parameter vector $\bfalpha$ was uniformly successful.  At all conditions, the bias and variability of the estimates were very small relative to the true values.  Note that accurate estimation of $\psi(\bfalpha^{\tT}\bfv)$  requires not only $\bfalpha$, but also $\bfgamma$ (the spline coefficients) to be estimated well.  We saw in panel (b) of Figure \ref{fig:Lambda0+psi+Linear} that moderate uncertainty about $\psi(\bfalpha^{\tT}\bfv)$ exists for $\bfalpha^{\tT}\bfv$ not in the vicinity of zero.  Given the quality of our $\bfalpha$ estimates, most of this uncertainty can be attributed to uncertainty in $\bfgamma$.  It is unclear whether this partitioning of the uncertainty about the nonlinear relationship---higher precision for relative effect sizes, lower precision for spline coefficients---is a general feature of the single index approach.  If it is, it would be an advantage of this method for the common situation where identifying important covariates is the primary research goal.  

For the linear regression coefficient $\beta$, the variances were reasonable, but the estimates exhibited some bias.  Interestingly, all estimates at $p=0.5$ were positively biased, while all but one at $p=1.5$ were negatively biased. It is possible that the $\beta$ estimate was compensating for some degree of model misspecification (for example, modelling the smooth baseline hazard functions by a four-piece step function). 

The association parameter $\phi$ proved to be the most difficult to estimate. It had large bias and large standard deviations, with estimation quality worsening as the true association gets weaker (larger $\phi$).  As with $\beta$, the bias is predominantly positive when $p=0.5$, and predominantly negative when $p=1.5$. For the $\phi=4$ cases, it was observed that A\{SE($\hat{\phi}$)\} and SD($\hat{\phi}$) were not very similar, indicating that the asymptotic assumptions probably do not hold at $\phi=4$ for our sample sizes. It is believed that the poor performance at this setting is due to the likelihood becoming very flat for large $\phi$ values.

A final curious observation from Table \ref{tab:othertable} is that in almost all cases, the bias in estimates of $\beta$ and $\phi$ decreased when the censoring rate was increased from 20\% to 50\%.  It appears that censoring introduces a bias that operates in the opposite direction to the biases in $\beta$ and $\phi$, causing the reduction.  The ``benefit'' of this new bias is offset by the considerable increase in variance that occurs when censoring increases.

\begin{table}
\caption{Estimation results for the factorial experiment.  The true values are $(\alpha_1, \alpha_2, \alpha_3, \beta)=(0.577, 0.577, 0.577, 1.0)$, and P(\%) refers to the censoring rate.}
\label{tab:othertable}\centering
\subfloat[$p=0.5$]{\label{tab:sim05}
\begin{scriptsize}
    \begin{tabular}{lcccr@{.}c@{ (}c@{)}cr@{.}c@{ (}c@{)}cr@{.}c@{ (}c@{)}cr@{.}c@{ (}c@{)}cr@{.}c@{ (}c@{)}}
    \hline\noalign{\smallskip}
    \multicolumn{3}{c}{Setting} && \multicolumn{19}{c}{Bias (SD)}\\ 
    $\phi$ & P($\%$) & $n$ &&  \multicolumn{3}{c}{$\alpha_1$} && \multicolumn{3}{c}{$\alpha_2$} && \multicolumn{3}{c}{$\alpha_3$} && \multicolumn{3}{c}{$\beta$} && \multicolumn{3}{c}{$\phi$}  \\\noalign{\smallskip}\hline\noalign{\smallskip}
    0.5    & 20\%    & 80  &&  -0 & 003 & 0.023             &&  0 & 000 & 0.024             &&  0 & 002 & 0.023                   && 0 & 171 & 0.205               &&  0 & 071 & 0.203              \\
           &         & 200 &&   0 & 001 & 0.012             && -0 & 001 & 0.013             &&  0 & 000 & 0.013                   && 0 & 168 & 0.128               &&  0 & 122 & 0.109              \\
           & 50\%    & 80  &&   0 & 000 & 0.054             && -0 & 003 & 0.056             && -0 & 005 & 0.048                   && 0 & 112 & 0.258               &&  0 & 019 & 0.318              \\
           &         & 200 &&   0 & 000 & 0.019             &&  0 & 000 & 0.018             && -0 & 001 & 0.019                   && 0 & 079 & 0.131               &&  0 & 029 & 0.142              \\
    1      & 20\%    & 80  &&  -0 & 001 & 0.026             && -0 & 002 & 0.026             &&  0 & 002 & 0.026                   && 0 & 171 & 0.216               &&  0 & 271 & 0.748              \\
           &         & 200 &&  -0 & 001 & 0.014             &&  0 & 001 & 0.015             &&  0 & 000 & 0.014                   && 0 & 164 & 0.120               &&  0 & 240 & 0.313              \\
           & 50\%    & 80  &&  -0 & 006 & 0.054             && -0 & 002 & 0.054             &&  0 & 001 & 0.048                   && 0 & 100 & 0.277               &&  0 & 063 & 0.816              \\
           &         & 200 &&  -0 & 003 & 0.022             && -0 & 002 & 0.022             &&  0 & 003 & 0.023                   && 0 & 083 & 0.162               &&  0 & 143 & 0.421              \\
    4      & 20\%    & 80  &&  -0 & 001 & 0.029             &&  0 & 002 & 0.025             && -0 & 003 & 0.028                   && 0 & 191 & 0.235               && -0 & 468 & 1.636              \\
           &         & 200 &&  -0 & 001 & 0.017             &&  0 & 001 & 0.018             && -0 & 001 & 0.016                   && 0 & 160 & 0.141               &&  0 & 682 & 1.768              \\
           & 50\%    & 80  &&   0 & 005 & 0.038             && -0 & 002 & 0.040             && -0 & 007 & 0.040                   && 0 & 107 & 0.301               && -1 & 493 & 1.579              \\
           &         & 200 &&  -0 & 004 & 0.023             &&  0 & 000 & 0.022             &&  0 & 003 & 0.025                   && 0 & 072 & 0.179               && -0 & 326 & 1.472              \\\noalign{\smallskip}\hline 
    \end{tabular}
\end{scriptsize}
}\\
\subfloat[$p=1.5$]{\label{tab:sim15}
\begin{scriptsize}
    \begin{tabular}{lcccr@{.}c@{ (}c@{)}cr@{.}c@{ (}c@{)}cr@{.}c@{ (}c@{)}cr@{.}c@{ (}c@{)}cr@{.}c@{ (}c@{)}}
    \hline\noalign{\smallskip}
    \multicolumn{3}{c}{Setting} && \multicolumn{19}{c}{Bias (SD)}\\
    $\phi$ & P($\%$) & $n$ &&  \multicolumn{3}{c}{$\alpha_1$} && \multicolumn{3}{c}{$\alpha_2$} && \multicolumn{3}{c}{$\alpha_3$} && \multicolumn{3}{c}{$\beta$} && \multicolumn{3}{c}{$\phi$}  \\\noalign{\smallskip}\hline\noalign{\smallskip}
    0.5    & 20\%    & 80  &&  -0 & 002 & 0.024             &&  0 & 003 & 0.023             && -0 & 002 & 0.023                   && -0 & 057 & 0.147               && -0 & 077 & 0.122              \\
           &         & 200 &&  -0 & 001 & 0.011             &&  0 & 001 & 0.011             &&  0 & 000 & 0.011                   && -0 & 074 & 0.081               && -0 & 063 & 0.070              \\
           & 50\%    & 80  &&   0 & 002 & 0.035             && -0 & 003 & 0.036             && -0 & 002 & 0.036                   && -0 & 035 & 0.231               && -0 & 091 & 0.238              \\
           &         & 200 &&  -0 & 002 & 0.019             &&  0 & 000 & 0.019             &&  0 & 001 & 0.019                   && -0 & 023 & 0.127               && -0 & 032 & 0.131              \\
    1      & 20\%    & 80  &&   0 & 000 & 0.027             && -0 & 005 & 0.024             &&  0 & 003 & 0.028                   && -0 & 039 & 0.176               && -0 & 159 & 0.429              \\
           &         & 200 &&   0 & 002 & 0.014             && -0 & 002 & 0.016             && -0 & 001 & 0.015                   && -0 & 056 & 0.102               && -0 & 132 & 0.196              \\
           & 50\%    & 80  &&  -0 & 001 & 0.038             && -0 & 001 & 0.039             && -0 & 001 & 0.037                   &&  0 & 002 & 0.240               && -0 & 102 & 0.702              \\
           &         & 200 &&   0 & 000 & 0.022             && -0 & 001 & 0.021             &&  0 & 001 & 0.019                   && -0 & 013 & 0.149               && -0 & 072 & 0.342              \\
    4      & 20\%    & 80  &&  -0 & 002 & 0.048             && -0 & 001 & 0.050             && -0 & 003 & 0.042                   && -0 & 030 & 0.194               && -1 & 642 & 1.208              \\
           &         & 200 &&  -0 & 002 & 0.018             && -0 & 001 & 0.017             &&  0 & 002 & 0.017                   && -0 & 046 & 0.115               && -0 & 474 & 1.526              \\
           & 50\%    & 80  &&   0 & 000 & 0.042             &&  0 & 003 & 0.042             && -0 & 007 & 0.043                   && -0 & 021 & 0.242               && -2 & 149 & 1.056              \\
           &         & 200 &&   0 & 001 & 0.024             && -0 & 005 & 0.023             &&  0 & 003 & 0.023                   && -0 & 020 & 0.157               && -0 & 774 & 1.377              \\
    \noalign{\smallskip}\hline 
    \end{tabular}
\end{scriptsize}
}
\end{table}

\section{Application to the Busselton Data}\label{Sec:Application}

The Busselton Health Study \citep{Knuiman1994} was a repeated cross-sectional survey that was conducted in the Busselton community in West Australia. From 1966 to 1981 a survey was conducted on adults in the community every three years. Various health-related information was collected, such as demographic variables, general health and lifestyle variables, health history variables, and physical, biochemical, haematological and immunological measurements.  We analyze these data using the proposed model, taking a subset of the available covariates, and using married couples as the clusters.

\subsection{Data description}\label{ssec:DataDesc}

The data set contains health information from 2306 couples  over 18 years old. The survival experience of the individuals, with survival time defined as age at death, is taken as our response. This is an example of parallel multivariate survival data, because the survival times of the husband and wife are likely to be associated, and there is no prior ordering associated with the times. The censoring rate is 80\% for females, and 67\% for males. Excluding the censored data, the average survival times for female and male are 75.2 and 74.2 respectively.

The data set has over ten health-related covariates.  We included four of them in our analysis.  Smoking status (SMOKE) was recorded as a binary code (1 to indicate a smoker, and 0 to indicate someone who has never smoked), so it was included as the linear covariate $x_{j}$. The other three covariates were age at the beginning of survey (AGE), body mass index (BMI), and total cholesterol (CHOL). These quantities, being continuous-valued, were were included in the nonlinear covariate vector $\bfv_j$. We standardized the covariates, to improve numerical stability during parameter estimation. The response survival times are the time to death ($t_{ij},~j=1,2$), where $j=1$ indicates wife, and $j=2$ denotes husband.

\subsection{Results}\label{ssec:Results}

Before starting to analyze the data using the proposed model, the assumption of proportional hazards was tested using the function \texttt{cox.zph} \citep{survival} in the R statistical computing environment \citep{R}.  It was found that the assumption is reasonable.

To find the interior cut points for the piecewise baseline hazard function estimates, we use the strategy of dividing the range into four intervals such that each interval has roughly the same survival probability according to the Kaplan-Meier curve. This procedure yields interior cut points of (78.1, 85.2, 90.9) for females and (74.4, 81.5, 87.4) for males. Similarly, the breakpoints used in the spline estimator of $\psi(\bfalpha^{\tT}\bfv)$ were chosen to divide the survival time range of each variate into four intervals, with each interval containing roughly an equal number of data points.

We use the same notation as before, with full parameter vector $(\phi, \bfrho^{\tT}, \bftau^{\tT}, \bfalpha^{\tT}, \beta, \bfgamma^{\tT})^{\tT}$.  In this case the baseline hazard functions have four constant pieces, so the parameter vectors $\bfrho$ (for females) and $\bftau$ (for males) have length four each. Parameter $\bfgamma$ has six elements for the six spline basis functions. Regression parameters $\alpha_1$, $\alpha_2$, $\alpha_3$, and $\beta$ are the coefficients for AGE, BMI, CHOL, and SMOKE, respectively.

Table \ref{tab:dataPH} gives the summary of the regression parameter estimates. Among the three continuous covariates, age has a dominating influence on survival time. The statistically significant covariates are AGE and SMOKE. The association parameter $\phi$ is also significant, with a point estimate of 5.12, which shows a mild degree of association between female and male.

\begin{table}
\caption{The estimates for the covariate coefficients $\bfalpha$, $\beta$ and $\phi$ in the Busselton data analysis}
\label{tab:dataPH}
\begin{center}
\begin{tabular}{lccccccc}
\hline\noalign{\smallskip}
& $\alpha_1$ & $\alpha_2$ & $\alpha_3$ & $\beta$ & $\phi$ \\
\noalign{\smallskip}\hline\noalign{\smallskip}
Estimate & 0.998 & 0.019 & 0.056 & 0.356 & 5.12 \\
SE($\hat{\cdot}$) & 0.004 & 0.049 & 0.064 & 0.076 & 2.14 \\
\noalign{\smallskip}\hline
\end{tabular}
\end{center}
\end{table}

\begin{figure*}
\centering
\subfloat[Baseline function (females)]{\label{fig:PHLam0female}\includegraphics[width=0.45\textwidth]{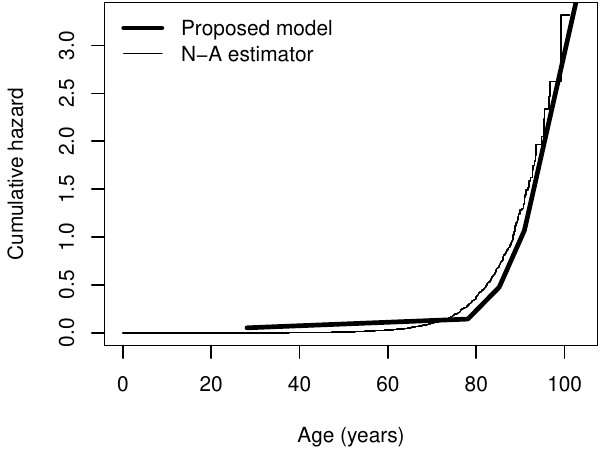}}
~~~\subfloat[Baseline function (males)]{\label{fig:PHLam0male}\includegraphics[width=0.45\textwidth]{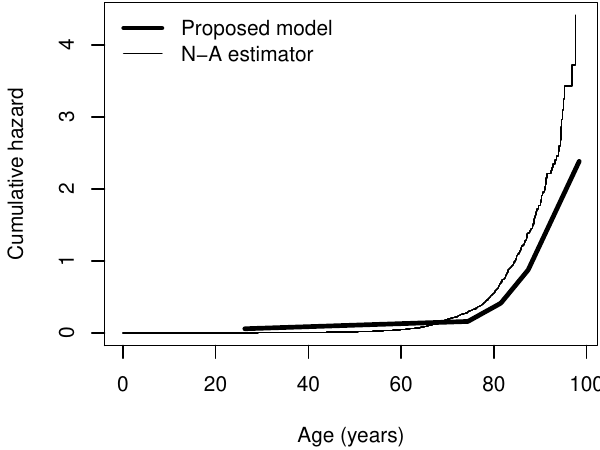}}\\
\subfloat[Individual curves (females)]{\label{fig:PHLamfemale}\includegraphics[width=0.45\textwidth]{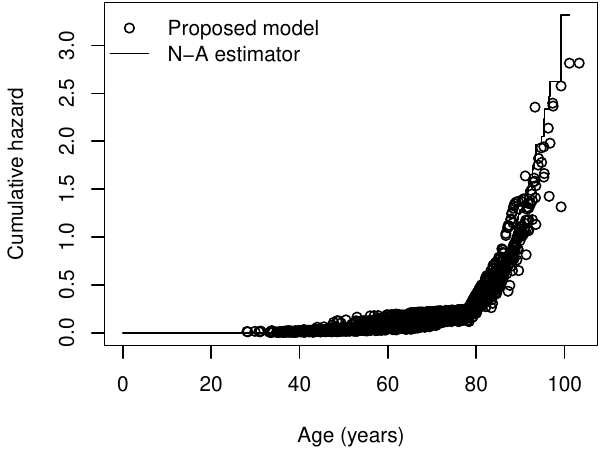}}
~~~\subfloat[Individual curves (males)]{\label{fig:PHLammale}\includegraphics[width=0.45\textwidth]{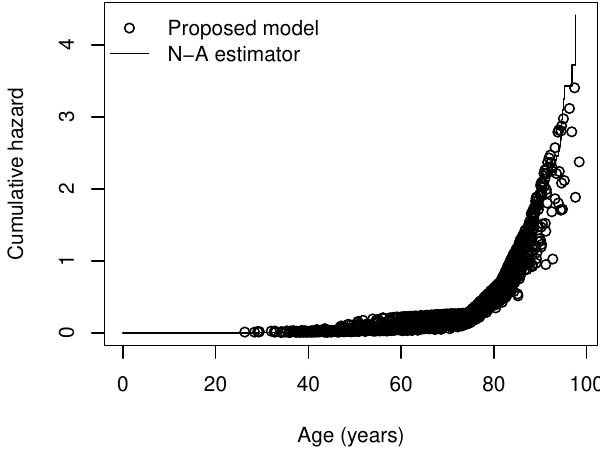}}
\caption{Comparison of the proposed model's cumulative hazard functions to the N-A estimator.}
\label{fig:PHLambda0}
\end{figure*}

The Nelson-Aalen (N-A) estimator \citep{N69JQT,A78AS} is a nonparametric estimator of the cumulative hazard function, computed from the survival data without considering covariate effects.  It can serve as a check to see if the proposed model's estimates are reasonable. Figure \ref{fig:PHLambda0} compares the proposed model's estimated cumulative hazards to the N-A estimates, separately for females and males.  The baseline functions are compared to the N-A estimates in the first row. The pairs of curves have similar general shape.  Differences between them is explained partly by the small number of pieces in the proposed model's estimates, and partly by the covariate effects that are not controlled for in the N-A estimator. The covariate effect is illustrated in the second row of plots, by showing points for each individual's estimated cumulative hazard (at their event time) on top of the N-A estimator.  It is easier to see in this figure that the proposed model's cumulative hazards are consistent with the N-A estimates. 

\begin{figure}
\centering
\label{fig:PHpsifemale}\includegraphics[width=0.55\textwidth]{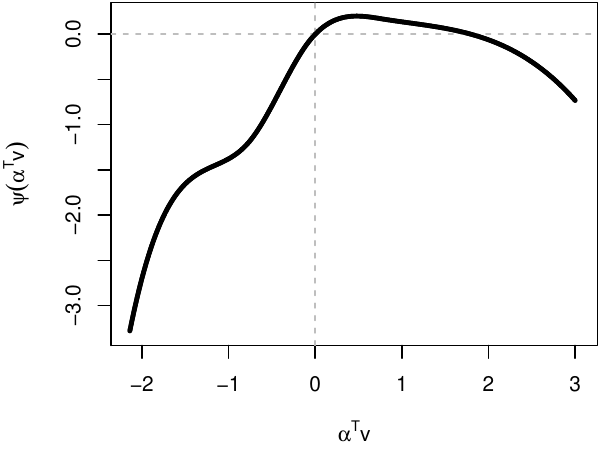}
\caption{Nonlinear function $\psi(\cdot)$.}
\label{fig:PHpsi}
\end{figure}

\begin{figure*}
\centering
\includegraphics[width=0.75\textwidth]{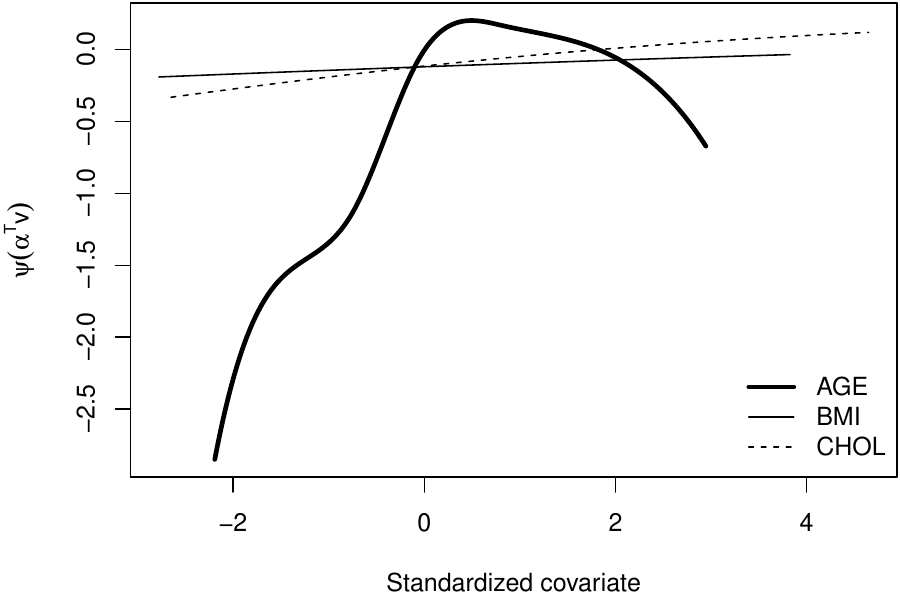}
\caption{The effect of single covariates on the shape of $\psi(\cdot)$, with the other covariates fixed at their median values. This plot is for males.}
\label{fig:PHpsiage}
\end{figure*}

Figure \ref{fig:PHpsi} shows the estimated single index function $\psi(\bfalpha^{\tT}\bfv)$ (which, under our assumptions, is the same for both males and females). The figure clearly shows a nonlinear relationship.  To better understand the effect of an individual covariate, we can plot $\psi(\bfalpha^{\tT}\bfv)$ as a function of that single covariate, while holding the other two covariates fixed at their median values.  This is shown, for all three covariates, in Figure \ref{fig:PHpsiage}. The plot is for male subjects only; the plot for females differs only slightly because of different median values, and is not shown.  It is clear that AGE is the dominant factor, both in terms of exhibiting nonlinearity and in its effect magnitude.  As age increases, $\psi(\cdot)$ experiences a steeper increase at the beginning, followed by a mildly decreasing stage that begins at roughly age 50. Therefore roughly from age 18 to 50 the death hazard is increasing, and after 50 it maintains a fairly high level. The BMI and CHOL effects are only slightly nonlinear, and both have a small positive effect on  $\psi(\cdot)$ (and the hazard) as they increase.  

The plots are consistent with the interpretation of the $\bfalpha$ parameters: $\alpha_1$ has by far the largest impact on mortality risk. If we had used the PH linear model, however, this would not be as clear: the normalized regression parameters for that model are $\hat{\bfalpha}=(0.856, 0.262, 0.446)$.  In addition to failing to reveal the nonlinearity of the age effect, the model would cause us to overestimate the relative importance of BMI\ and CHOL.  

\section{Summary and Discussion}\label{sec:Summary}

In this work, the PH model was extended by adding a nonlinear covariate effect to the log hazards ratio through a single index structure.  Clustered data was handled through a copula model.  Through appropriate use of weakly-parametric elements, we were able to formulate the the full likelihood and carry out maximum likelihood  inference in the standard manner. Major advantages of this model are the ability to flexibly model the covariate effect, including nonlinear relationships, and the ability to calculate the survival function and hazard function for each individual. 

Through simulation, it was found that the proposed model is able to capture nonlinear relationships well.  This is a clear advange for cases where covariate effects are truly nonlinear. In the analysis of the Busselton data, mild association was found to exist between husbands and wives. The hazard function considering the covariate effects could be estimated for each individual. Nonlinear covariate effects were clearly visible in these data, particularly for the age at enrollment.  Capturing the nonlinearity of the age effect allowed a more accurate interpretation of all the covariates and their relative importance.   

The price paid for the convenience of full likelihood inference is the number of parameters in the model.  The Busselton data analysis, for example, was a relatively small example with only four covariates, four pieces for each baseline function, and six spline coefficients.  Including the association parameter, 19 parameters in all were estimated.  If we had tens of covariates, increased the number of knots and cut points to get more flexible weakly-parametric estimates, and allowed separate single-index functions for each member of a cluster, the parameter count would grow quickly.  This problem would be further exacerbated if we attempted to model clusters larger than two.  While we found our optimization problems to be numerically well-behaved after the parameter transformations, there is room for future work on the computational aspects of models like this, including numerical studies of the trade-off between model complexity and estimation efficiency for different sample sizes and degrees of nonlinearity.  

The extensions demonstrated in this paper could also be applied to other survival models.  Of particular interest are the proportional odds model and the generalized transformation model \citep[see, e.g.,][]{Cheng1995}.  We have promising initial results on these models and plan to study the relative benefits of the proportional hazards, proportional odds, and transformation models in the case of clustered data with nonlinear covariate effects. 


\bibliographystyle{spbasic}      
\bibliography{bibdata-clean} 
%
%

\end{document}